\documentclass[aps, pra,superscriptaddress,showpacs,twocolumn,10pt]{revtex4-1}

\usepackage{amsfonts,amssymb,amsmath}
\usepackage{graphics,graphicx,epsfig}
\usepackage{amsthm}
\usepackage{epstopdf}

\newcommand{\ket}[1]{ | #1 \rangle}
\newcommand{\bra}[1]{ \langle #1  |}

\newcommand{\Tr}{\mathrm{Tr}}

\newcommand{\proj}[1]{\ket{#1}\bra{#1}}
\renewcommand{\epsilon}{\varepsilon}

\def\identity{\leavevmode\hbox{\small1\kern-3.8pt\normalsize1}}

\begin{document}

\title{Genuine Multipartite Entanglement without Multipartite Correlations}

\author{Christian~Schwemmer}
\affiliation{Max-Planck-Institut f\"ur Quantenoptik, Hans-Kopfermann-Stra{\ss}e 1, 85748 Garching, Germany}
\affiliation{Department f\"ur Physik, Ludwig-Maximilians-Universit\"at, 80797 M\"unchen, Germany}

\author{Lukas~Knips}
\affiliation{Max-Planck-Institut f\"ur Quantenoptik, Hans-Kopfermann-Stra{\ss}e 1, 85748 Garching, Germany}
\affiliation{Department f\"ur Physik, Ludwig-Maximilians-Universit\"at, 80797 M\"unchen, Germany}

\author{Minh Cong Tran}
\affiliation{School of Physical and Mathematical Sciences, Nanyang Technological University, 637371 Singapore}

\author{Anna~de~Rosier}
\affiliation{Institute of Theoretical Physics and Astrophysics, University of Gda\'nsk, PL-80-952 Gda\'nsk, Poland}

\author{Wies{\l}aw~Laskowski}
\affiliation{Institute of Theoretical Physics and Astrophysics, University of Gda\'nsk, PL-80-952 Gda\'nsk, Poland}

\author{Tomasz~Paterek}
\email{tomasz.paterek@ntu.edu.sg}
\affiliation{School of Physical and Mathematical Sciences, Nanyang Technological University, 637371 Singapore}
\affiliation{Centre for Quantum Technologies, National University of Singapore, 117543 Singapore}
\affiliation{MajuLab, CNRS-UNS-NUS-NTU International Joint Research Unit, UMI 3654, Singapore}

\author{Harald~Weinfurter}
\affiliation{Max-Planck-Institut f\"ur Quantenoptik, Hans-Kopfermann-Stra{\ss}e 1, 85748 Garching, Germany}
\affiliation{Department f\"ur Physik, Ludwig-Maximilians-Universit\"at, 80797 M\"unchen, Germany}

\begin{abstract}
Non-classical correlations between measurement results make entanglement the essence of quantum physics and the main resource for quantum information applications. Surprisingly, there are $n$-particle states which do not exhibit $n$-partite correlations at all but still are genuinely $n$-partite entangled. We introduce a general construction principle for such states, implement them in a multiphoton experiment and analyze their properties in detail. Remarkably, even without $n$-partite correlations, these states do violate Bell inequalities showing that there is no classical, i.e., local realistic model describing their properties.
\end{abstract}


\date{\today}
\maketitle

Correlations between measurement results are the most prominent feature of entanglement.
They made Einstein, Podolski and Rosen \cite{PhysRev.47.777} to question the completeness of quantum mechanics, and are nowadays the main ingredient for the many applications of quantum information like entanglement based quantum key distribution \cite{PhysRevLett.67.661} or quantum teleportation \cite{PhysRevLett.70.1895}.

Correlations enable us, e.g., when observing two maximally entangled qubits, to use a measurement result observed on the first system to infer exactly the measurement result on the second system. 
In this scenario the two particle correlations are formally given by the expectation value of the product of the measurement results obtained by the two observers. Note, the single particle correlation, i.e., the expectation value of the results for one or the other particle are zero in this case. Consequently, we cannot predict anything about the individual results.
When studying the entanglement between $n$ particles, a natural extension is to consider $n$-partite correlations, i.e., the expectation value of the product of $n$ measurement results. Such correlation functions are frequently used in classical statistics and signal analysis \cite{statistics}, moreover in quantum information almost all standard tools for analyzing $n$-partite systems like multi-party entanglement witnesses \cite{witnesses,entanglementcriterion} and Bell inequalities \cite{bellinequalities,PhysRevLett.88.210401} are based on the $n$-partite correlation functions. 

Recently, Kaszlikowski \emph{et al.}~\cite{PhysRevLett.101.070502} pointed at a particular quantum state with vanishing multi-party correlations which, however, is genuinely multipartite entangled. This discovery, of course, prompted vivid discussions on a viable definition of classical and quantum correlations \cite{correlations,PhysRevA.83.012312}.
Still, the question remains what makes up such states with no full $n$-partite correlations and how non-classical they can be, i.e., whether they are not only entangled but whether they also violate a Bell inequality.

%

Here, we generalize, highlight and experimentally test such remarkable quantum states.
We introduce a simple principle how to construct states without $n$-partite correlations for odd $n$ and show that there are infinitely many such states which are genuinely $n$-partite entangled. 
We implement three and five qubit no-correlation states in a multiphoton experiment and demonstrate that these states do not exhibit $n$-partite correlations.
Yet, due to the existence of correlations between a smaller number of particles, we observe genuine $n$-partite entanglement. 
Using our recently developed method to design $n$-partite Bell inequalities from lower order correlation functions only~\cite{PhysRevA.86.032105,PhysRevA.82.012118}, we show that these states, despite not having full correlations, can violate Bell inequalities.

\emph{Correlations.}---The quantum mechanical correlation function $T_{j_1 \dots j_n}$ is defined as the expectation value of the product of the results of $n$ observers
\begin{equation}
T_{j_1 \dots j_n} = \langle r_1 \dots r_n \rangle = \Tr(\rho~ \sigma_{j_1} \otimes \dots \otimes \sigma_{j_n}),
\label{CORR}
\end{equation}
where $r_k$ is the outcome of the local measurement of the $k$-th observer, parametrized by the Pauli operator $\sigma_{j_k}$ with $j_k\in\{x,y,z\}$.
Evidently, besides the $n$-partite correlations, for an $n$-party state one can also define $l<n$ fold correlations  $T_{\mu_1 \dots \mu_{n}} = \Tr(\rho ~\sigma_{\mu_1} \otimes \dots \otimes \sigma_{\mu_{n}})$ with $\mu_i\in\{0,x,y,z\}$ and $|\{\mu_i = 0\}| = n-l$.
Non-vanishing $l$-fold correlations indicate that we can infer (with higher probability of success than pure guessing) an $l$-th measurement result from the {\em product} of the other $(l-1)$ results (see Supplemental Material~(SM)~(see Appendix \ref{A})). Only in the two particle scenario we can directly use the result from one measurement to infer the other result.
For an $n$-qubit no-correlation state the vanishing $n$-partite correlations do not imply vanishing correlations between a smaller number of observers, thus not necessarily destroying predictability. We will see also in the experimentally implemented example that the various individual results still enable some possibility for inference, which is then largely due to bipartite correlations.


\emph{Constructing no-correlation states.}---For any state $\ket{\psi}$ with an odd number $n$ of qubits we can construct an ``anti-state'' $\ket{\overline{\psi}}$, i.e., the state whose $n$-partite correlations are inverted with respect to the initial one. By evenly mixing these states
\begin{equation}
\rho^{nc}_{\psi} = \frac{1}{2} \proj{\psi} + \frac{1}{2} \proj{\overline{\psi}},
\label{E_MIXED}
\end{equation}
we obtain a state $\rho^{nc}_{\psi}$ without $n$-partite correlations.

The anti-state $| \overline{\psi} \rangle$ of a state $\ket{\psi}$ described in the computational basis by
\begin{equation}
\ket{\psi} = \sum_{k_1, \dots, k_n = 0}^1 \alpha_{k_1, \dots, k_n} \ket{k_1 \dots k_n},
\end{equation}
with normalized coefficients $\alpha_{k_1, \dots, k_n} \in  \mathbb{C}$, is given by
\begin{equation}
| \overline{\psi} \rangle \equiv \sum_{k_1, \dots, k_n = 0}^1 (-1)^{k_1 + \dots + k_n} \alpha_{1 - k_1, \dots, 1- k_n}^* \ket{k_1 \dots k_n}, \end{equation}
where the asterisk denotes complex conjugation. This state has inverted correlations with respect to those in $\ket{\psi}$ for every odd number of observers, whereas all the correlation function values for an even number of observers remain unchanged. 

$| \overline{\psi} \rangle$ is mathematically obtained from $\ket{\psi}$ by applying local universal-not gates~\cite{JModOpt.47.211}. These gates introduce a minus sign to all local Pauli operators.
Therefore, for odd $n$ the correlations of $|\overline{\psi} \rangle$ have opposite sign to those of $\ket{\psi}$.
Representing the universal-not gate by $N = \sigma_z \sigma_x K$, where $K$ is the complex conjugation operating in the computational basis, i.e. $K (\alpha \ket{0} + \beta \ket{1}) = \alpha^* \ket{0} + \beta^* \ket{1}$, indeed, we obtain  $N\sigma_x N^\dagger = -\sigma_x$, $N \sigma_y N^\dagger = -\sigma_y$, and $N \sigma_z N^\dagger = -\sigma_z$.
Applying $N$ to all the $n$ subsystems we find the anticipated result $N \otimes \dots \otimes N \ket{\psi} = | \overline{\psi} \rangle$.

Although $N$ is antiunitary, $| \overline{\psi} \rangle$ is always a proper physical state and can be obtained by some global transformation of $\ket{\psi}$. In general, $N$ can be approximated~\cite{universalnotgate}, but if all the coefficients $\alpha_{k_1 \dots k_n}$ are real, complex conjugation can be omitted and no-correlation states can be generated by local operations.

This construction principle can be generalized to mixed states using $\overline{\rho} = N^{\otimes n} \rho \left(N^{\otimes n}\right)^\dagger$, which changes every pure state in the spectral form to the respective anti-state. Evenly mixing $\rho$ and $\overline{\rho}$ therefore produces a state with no $l$-party correlations for all odd $l$.

One may then wonder whether the principle of Eq.~(\ref{E_MIXED}) can also be applied to construct a no-correlation state for every state with an even number of qubits.
The answer is negative as shown by the following counterexample.
Consider the Greenberger-Horne-Zeilinger state of an even number of qubits $\ket{\psi} = \frac{1}{\sqrt{2}}(\ket{0\dots0} + \ket{1\dots1})$.
It has non-vanishing $T_{z \dots z}$, $2^{n-1}$ $n$-partite correlations in the $xy$-plane, and also $2^{n-1}-1$ correlations between a smaller number of subsystems, all equal to $\pm 1$.
However, for a state with inverted correlations between all $n$ parties (making no assumptions about the correlations between smaller numbers of observers) the fidelity relative to the GHZ state, given by $\frac{1}{2^n} \sum_{\mu_1, \dots, \mu_n = 0}^3 T_{\mu_1 \dots \mu_n}^{\mathrm{GHZ}} T_{\mu_1 \dots \mu_n}^{\textrm{anti}}$,
is negative because more than half of the correlations are opposite. Hence this state is unphysical and there is no such ``anti-state''.
In fact, so far we were unable to find an anti-state to \emph{any} genuinely multi-qubit entangled state of even $n$.

\emph{Entanglement without correlations: infinite family.}---Consider a three-qubit system in the pure state 
\begin{equation}
\ket{\phi} = \sin \beta \cos \alpha \ket{001} + \sin \beta \sin \alpha \ket{010} + \cos \beta \ket{100},
\label{GEN_W}
\end{equation}
where  $\alpha, \beta \in (0, \frac{\pi}{2})$ (which includes the state $\ket{W}$ with $\alpha=\pi/3$ and $\beta=\cos^{-1}(1/\sqrt{3})$).
Together with any local unitary transformation thereof this defines a three dimensional subspace of genuinely tripartite entangled states within the eight dimensional space of three qubit states.
To show that all the respective no-correlation states $\rho^{nc}_{\phi}$ are genuinely entangled, we use a criterion similar to the one in \cite{entanglementcriterion}, i.e. 
\begin{equation}
\max_{T^{\mathrm{bi-prod}}} (T,T^{\mathrm{bi-prod}}) < (T,T^{\mathrm{exp}}) \Rightarrow \rho^{\mathrm{exp}} \textrm{ is not bi-sep},
\label{COND}
\end{equation}
where maximization is over all bi-product pure states and $(U,V) \equiv \sum_{\mu,\nu,\eta=0}^3 U_{\mu \nu \eta} V_{\mu \nu \eta}$ denotes the inner product in the vector space of correlation tensors. 
Condition (\ref{COND}) can be interpreted as an entanglement witness
$\mathcal{W} = \alpha \openone - \rho^{nc}_{\phi}$, where $\alpha = L/8$ and $L=\max_{T^{\mathrm{bi-prod}}} (T,T^{\mathrm{bi-prod}})$ is the left-hand side of (\ref{COND}). 
In the ideal case of preparing $\rho^{\mathrm{exp}}$ perfectly, $T^{\mathrm{exp}} = T$, the right-hand side of our criterion equals $4$ for all the states of the family, and thus the expectation value of the witness is given by ${\rm Tr}(\mathcal{W} \rho^{nc}_{\phi}) = (L-4)/8.$ 

A simple argument for $\rho^{nc}_{\phi}$ being genuinely tripartite entangled can be obtained from the observation that $\ket{\phi}$ and $\ket{\overline{\phi}}$ span a two-dimensional subspace of the three qubit Hilbert space~\cite{PhysRevLett.101.070502}. 
As none of the states $\ket{\Phi} = a \ket{\phi} + b\ket{\overline{\phi}}$ is biproduct (for the proof see SM~(see Appendix \ref{B})), states in their convex hull do not intersect with the subspace of bi-separable states and thus all its states, including $\rho^{nc}_{\phi}$ are genuinely tripartite entangled. 
To evaluate the entanglement in the experiment, we calculated $L$ for all states of Eq. (\ref{GEN_W}). We obtain $L_{\ket{\phi}}<4$ in general, with $L_{\ket{W}}=10/3$. Similar techniques were used to analyze five-qubit systems.


\emph{Quantum correlations without classical correlations?}---The cumulants and correlations were initially proposed as a measure of genuinely multi-party non-classicality in Ref.~\cite{PhysRevA.74.052110}. Kaszlikowski {\em et al.} \cite{PhysRevLett.101.070502}, however, showed that such a quantification is not sufficient as the state $\rho_{W}^{nc}$ has vanishing cumulants, yet contains genuinely multi-party entanglement.
They suggested that the vanishing cumulants or standard correlation functions (\ref{CORR}) indicate the lack of genuine multi-party ``classical'' correlations.
This initiated a vivid discussion on a proper definition and measure of genuine multipartite ``classical'' and quantum correlations. Bennett {\em et al.} proposed a set of axioms for measures of genuine multipartite correlations~\cite{PhysRevA.83.012312}. They showed that the correlation function (\ref{CORR}) does not fulfill all the requirements, but also still strive for computable measures that satisfy these axioms~\cite{PhysRevLett.107.190501,QuantInfProc.12.2371}. An information-theoretic definition of multipartite correlations was given by Giorgi \emph{et al.}~\cite{PhysRevLett.107.190501}.
Their measure combines the entropy of all sizes of subsystems. 
Applying their definitions to $\rho^{nc}_{W}$, we obtain genuine classical tripartite correlations of 0.813 bit and genuine quantum tripartite correlations of 0.439 bit resulting in total genuine tripartite correlations of 1.252 bit (see SM~(see Appendix \ref{C}) for calculations for all $\rho^{nc}_{\phi}$). While this approach does assign classical correlations in the context of Giorgi \emph{et al.}~\cite{PhysRevLett.107.190501} to $\rho_{W}^{nc}$, it does not fulfill all requirements of \cite{PhysRevA.83.012312} either.


\begin{figure}
\includegraphics[width=0.40\textwidth]{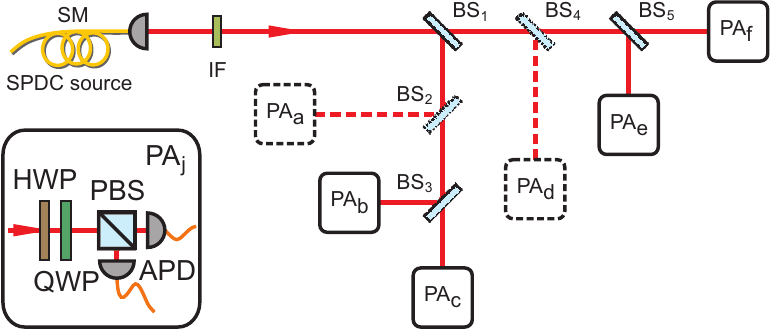}
\caption{(color online). Schematic of the linear optical setup used to observe symmetric Dicke states from which states with vanishing $3$- and $5$-partite correlations can be obtained. The photons are created by means of a cavity enhanced pulsed collinear type II spontaneous parametric down conversion source pumped at $390$nm \cite{NaturePhotonics.4.170}. Distributing the photons symmetrically into six modes by 5 beam splitters (BS) enables the observation
of the state $\vert D_{6}^{(3)}  \rangle$. Removing beam splitters BS$_{2}$ and BS$_{4}$ reduces the number of modes to four and thus the state $\vert D_{4}^{(2)}  \rangle$ is obtained.
State analysis is enabled by sets of half-wave (HWP) and quarter-wave plates (QWP) together with polarizing beam splitters (PBS) in each mode.
The photons are measured by fiber-coupled single photon counting modules connected to a coincidence logic~\cite{setups}.}
\label{FIG_SETUP}
\end{figure}

\emph{Experiment.}---The three photon state $\ket{W}$ can be observed either using a multiphoton interferometer set-up~\cite{PhysRevLett.92.077901} or by suitably projecting the fourth photon of a $4$-photon symmetric Dicke state~\cite{PhysRevA.66.064301}.
The latter scheme has the advantage that it also offers the option to prepare the states $\ket{\overline{W}}$ and $\rho^{nc}_W$.
The states $\ket{W}$ and $\ket{\overline{W}}$ are particular representatives of the symmetric Dicke states, which are defined as
\begin{eqnarray}
\vert D_{n}^{(e)}  \rangle =\binom{n}{e}^{-1/2} \sum_i \mathcal{P}_i(\vert H^{\otimes(n-e)} \rangle \otimes  \vert V^{\otimes e} \rangle )
\end{eqnarray}
where $\ket{H/V}$ denotes horizontal/vertical polarization and $\mathcal{P}_i$ all distinct permutations, and with the three photon states $\ket{W}=\vert D_{3}^{(1)}$ and $\ket{\overline{W}}=\vert D_{3}^{(2)}$.
We observed four- and six-photon Dicke states using a pulsed collinear type II spontaneous parametric down conversion source together with a 
linear optical setup (see Fig. \ref{FIG_SETUP}) ~\cite{setups, NaturePhotonics.4.170}. 
The $| D_{n}^{(e)} \rangle$ states were observed upon detection of one photon in each of the four or six spatial modes, respectively.
We characterized the state $| D_{4}^{(2)} \rangle$ by means of quantum state tomography, i.e., a polarization analysis in each mode, collecting for each setting 26 minutes of data at a  rate of 70 events per minute. The fidelity of the experimental state $| D_4^{(2)} \rangle^{exp}$ was directly determined from the observed frequencies together with Gaussian error propagation as $0.920 \pm 0.005$, which due to the high number of detected events~\cite{arxiv} is compatible with the value $0.917 \pm 0.002$ as obtained from a maximum likelihood (ML) reconstruction and non-parametric bootstrapping (see Appendix \ref{E}, \cite{bootstrapping}).
The high quality achieved here allowed a precise study of the respective states. 
The fidelities of the observed three qubit states with respect to their target states are  $0.939\pm0.011$ for $\ket{W}^{exp}$, $0.919\pm0.010$ for $\ket{\overline{W}}^{exp}$, and $0.961\pm0.003$ for $\rho^{nc,exp}_W$. 
Analogously, starting with a six-photon Dicke state $| D_6^{(3)} \rangle$ \cite{PhysRevLett.103.020504} 
we could also analyze the properties of the five photon state
$\rho^{nc}_{D_{5}^{(2)}}$.
The five-qubit fidelity of $\rho^{nc,exp}_{D_{5}^{(2)}}$ is determined via a ML reconstruction from five-fold coincidences to be $0.911\pm0.004$ (for the detailed characterization see SM~(see Appendix \ref{D})).

\begin{figure}
 \includegraphics[width=0.42\textwidth]{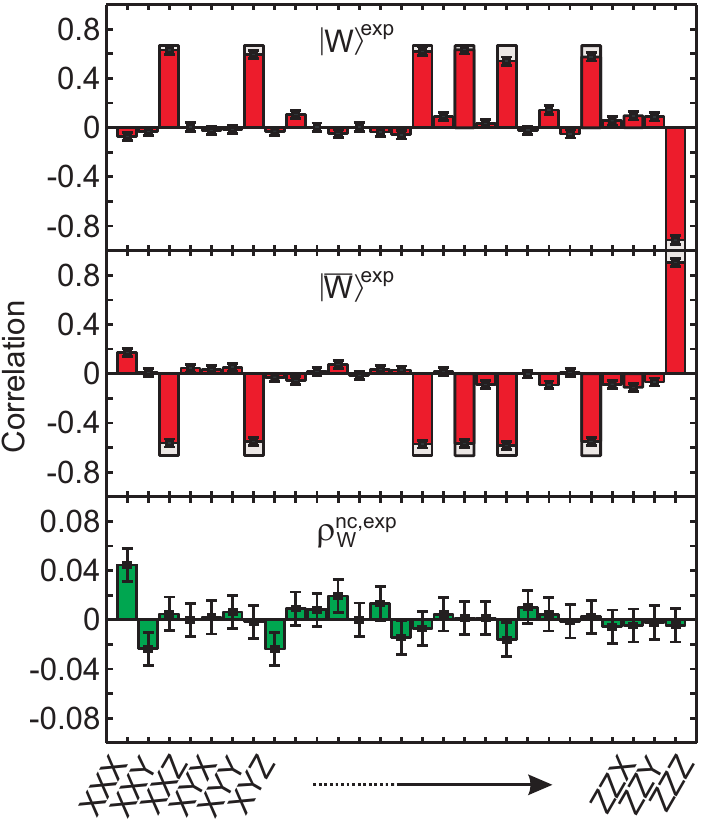}
    \caption{(color online). Experimental tripartite correlations (red) for $\ket{W}^{exp}$, $\ket{\overline{W}}^{exp}$, and (green) $\rho_W^{nc,exp}$ in comparison to the theoretically expected values (gray).    
    Note that the correlations of the state $\rho_W^{nc,exp}$ are magnified by a factor of 10. 
    The plot presents measured values of $T_{j_1 j_2 j_3}$ for the observables listed below the plot. Obviously, the states $\ket{W}^{exp}$ and $\ket{\overline{W}}^{exp}$ have opposite tripartite correlations canceling each other when they are mixed. 
}    
    \label{FIG_NOCORR}
\end{figure}
\begin{figure}[!b]
\includegraphics[width=0.46\textwidth]{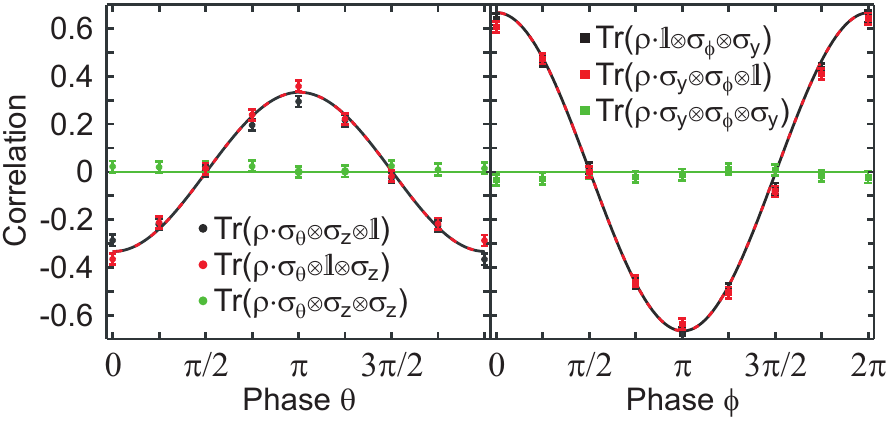}
\caption{(color online). Vanishing tripartite correlations for arbitrary measurements and non-vanishing bipartite correlations. Observable $\sigma_{\theta}$ ($\sigma_{\phi}$) was measured on the first (second) qubit and $\sigma_z$ ($\sigma_y$) measurements were performed on both other qubits (green curves) or one of them (red and black curves). The solid lines show the theoretically expected curves.}
\label{FIG_ROT}
\end{figure}
For the experimental analysis of the states, we start by determining $T_{zzz}$ for the three states $\ket{W}^{exp}$, $\ket{\overline{W}}^{exp}$, and $\rho^{nc,exp}_W$. As the first two have complementary structure of detection probabilities (with $T_{zzz}=-0.914\pm0.034$ and $T_{zzz}=0.904\pm0.034$, respectively), weighted mixing of these states leads to $\rho^{nc,exp}_W$ with $T_{zzz}=0.022\pm 0.023 $, i.e. a correlation value compatible with $0$ (see SM). 
Fig.~\ref{FIG_NOCORR} presents experimental data for all possible tripartite correlations of the observed states. 
Assuming a normal distribution centered at zero with a standard deviation given by our experimental errors, the observed correlations have a $p$-value of $0.44$ for the Anderson-Darling 
test, which shows that indeed one can adhere to the hypothesis of vanishing full correlations.
Similarly, the five qubit state $\rho^{nc,exp}_{D_{5}^{(2)}}$ exhibits strongly suppressed, almost vanishing correlations. 
For details on the five qubit state, please see SM~(see Appendix \ref{D}). 

We want to emphasize that the vanishing tripartite correlations of $\rho_W^{nc,exp}$ are no artifact of measuring in the Pauli bases.
In fact, all states obtained via local unitary transformations do not exhibit any $n$-partite correlations. To illustrate this property, 
we considered correlation measurements in non-standard bases. 
As an example, we chose measurements in the $zy$-plane $\sigma_{\theta} = \cos \theta \, \sigma_z + \sin \theta \, \sigma_y$ with $\theta \in [0,2\pi]$ ($\sigma_{\phi} = \cos \phi \, \sigma_y + \sin \phi \, \sigma_z$ with $\phi \in [0,2\pi]$) for the first (second) qubit resulting in the correlations $T_{\theta\,j_2\,j_3} = {\rm Tr}(\rho ~\sigma_{\theta} \otimes \sigma_{j_2} \otimes  \sigma_{j_3})$ ($T_{j_1\,\phi\,j_3}$).
Indeed, as shown in Fig.~\ref{FIG_ROT}, $T_{\theta\,j_2\,j_3}$ ($T_{j_1\,\phi\,j_3}$) vanishes independently of the choice of $\theta$ ($\phi$).
In contrast, the bipartite correlations $T_{\theta\,z0}$ ($T_{y\,\phi\,0}$) between qubit $1$ and $2$ do not vanish at all and clearly depend on $\theta$ ($\phi$). By means of those even number correlations, one is still able to infer the result of another party from ones own result with probability $2/3>1/2$. 
For example, the values of $T_{zz0}=-1/3$ ($T_{z0z}=-1/3$) indicate that knowing, e.g., result ``0'' for the first qubit, we can infer that the result will be ``1'' with $p=2/3$ on the second (third) qubit, etc.

Although the three qubits are not tripartite correlated, the \emph{bi}partite correlations shown above give rise to genuine \emph{tri}partite entanglement.
This can be tested for the experimental states employing (\ref{COND}). We observe 
\begin{eqnarray}
(T, T_W^{nc,exp}) & = & 3.858 \pm 0.079 > 3.33\overline{3}, \nonumber \\
(T, T_{D_5^{(2)}}^{nc,exp}) & = & 13.663 \pm 0.340 > 12.8, \nonumber
\end{eqnarray}
both above the respective bi-separable bound of $10/3 = 3.33\overline{3}$ ($12.8$) by more than $6.6$ ($2.4$) standard deviations, proving that in spite of vanishing full correlations the  states are genuinely tripartite (five-partite) entangled~(see Appendix \ref{E}).

The observed five-photon state has one more remarkable property~\cite{PhysRevA.86.032105}.
For this state, every correlation between a fixed number of observers, i.e., bipartite correlations, tripartite correlations, etc. admits description with an explicit local hidden-variable model \cite{PhysRevLett.88.210401}. 
However, some of the models are different and thus cannot be combined in a single one.
Using linear programming to find joint probability distributions reproducing quantum predictions~\cite{PhysRevA.82.012118}, we obtain an optimal Bell inequality using only two- and four-partite correlations~\cite{PhysRevA.86.032105}.
From the observed data we evaluate the Bell parameter to be $\mathcal{B} = 6.358 \pm 0.149$ which violates the local realistic bound of $6$ by $2.4$ standard deviations~\cite{footnote}. 
This violation confirms the non-classicality~(see Appendix \ref{E}) of this no-correlation state and also offers its applicability for quantum communication complexity tasks.
Contrary to previous schemes, here, the communication problem can be solved in every instance already by only a subset of the communicating parties~\cite{gdansk367}.


\emph{Conclusions.}---We introduced a systematic way to define and to experimentally observe mixed multipartite states with no $n$-partite correlations for odd $n$, as measured by standard correlation functions. 
For the first time we experimentally observed a state which allowed the violation of a Bell inequality without full correlations,  thereby proving both the  non-classicality of no-correlation states as well as their applicability for quantum communication protocols.
The remarkable properties of these states prompt intriguing questions. 
For example, what might be the dimensionality of these states or their respective subspaces, or whether we can even extend the subspace of states and anti-states which give genuinely entangled no-correlation states? 
Moreover, can no-correlation states be used for quantum protocols beyond communication complexity, 
and, of course, whether these remarkable features can be cast into rigorous and easily calculable measures of genuine correlations satisfying natural postulates~\cite{PhysRevA.83.012312}?

\begin{acknowledgments}

\emph{Acknowledgments.}---We thank the EU-BMBF project QUASAR and the EU projects QWAD and QOLAPS for supporting this work.
TP acknowledges support by the National Research Foundation, the Ministry of Education of Singapore grant no. RG98/13, start-up grant of the Nanyang Technological University, and NCN Grant No. 2012/05/E/ST2/02352. CS and LK thank the Elite Network of Bavaria for support (PhD programs QCCC and ExQM).

\end{acknowledgments}

\appendix

\section{Physical meaning of correlation functions}

\label{A}

Correlations for two particles are often seen as a measure of predictability of local results when knowing the other result.
Yet, this simple statement has to be used carefully.
A non-vanishing $n$-partite correlation function indicates that we can make an educated guess of the $n$th result from the product of the other $n-1$ results.
The converse statement does not hold and we provide an example of a state with vanishing correlation functions where the inference is still possible.

Let us denote by $r_j = \pm 1$ the result of the $j$th observer.
We assume that $n-1$ parties cannot infer from the product of their outcomes, $r_1 \dots r_{n-1}$, the result of the last observer, $r_n$, i.e., the following conditional probabilities hold:
\begin{equation}
P(r_n | r_1 \dots r_{n-1}) = \frac{1}{2}.
\label{COND_PROB}
\end{equation}
We show that this implies that the corresponding correlation function, $T_{j_1 \dots j_n}$, vanishes.
The correlation function is defined as expectation value of the product of all local outcomes
\begin{equation}
T_{j_1 \dots j_n} = \langle r_1 \dots r_n \rangle = P(r_1 \dots r_n = 1) - P(r_1 \dots r_n = -1).
\end{equation}
Using Bayes' rule
\begin{eqnarray}
P(r_1 \dots r_n = \pm 1) &=& \sum_{r = \pm 1} P(r_n = \pm r | r_1 \dots r_{n-1} = r) \nonumber \\
&\times& P(r_1 \dots r_{n-1} = r). 
\end{eqnarray}
According to assumption (\ref{COND_PROB}) we have $P(r_n = \pm r | r_1 \dots r_{n-1} = r) = \frac{1}{2}$, giving $P(r_1 \dots r_n = \pm 1) = \frac{1}{2}$ and $T_{j_1 \dots j_n} = 0$.

As an example of a state with vanishing correlation functions yet allowing to make an educated guess of the result, let us consider the two-qubit mixed state
\begin{equation}
\frac{1}{2} \proj{00} + \frac{1}{4} \proj{01} + \frac{1}{4} \proj{10},
\end{equation}
where $\ket{0}$ and $\ket{1}$ are the eigenstates of the Pauli operator $\sigma_z$ with eigenvalues $+1$ and $-1$, respectively.
All correlation functions $T_{kl}$, with $k,l=x,y,z$, of this state vanish. Yet, whenever Alice (Bob) observes outcome $-1$ in the $\sigma_z$ measurement, she (he) is sure the distant outcome is $+1$, i.e., $P(r_2 = +1 | r_1 = -1) = 1$.
Similar examples exist for multiple qubits, but we note that the states $\rho^{nc}_{\phi}$ of the main text are an equal mixture of a state and its anti-state. In this case, the vanishing $n$-party correlations lead to the impossibility of inferring the $n$-th result.

\section{Criterion for genuine multipartite entanglement}

\label{B}

To evaluate entanglement we use the following criterion (see main text) where, $T^{{exp}} = T$, i.e., assuming the ideal experiment producing the required state described by the correlation tensor $T$:
\begin{equation}
\max_{T^{{bi-prod}}} (T,T^{{bi-prod}}) < (T,T).
\end{equation}
The maximization is performed over all bi-product states keeping in mind also all possible bipartitions.
The inner product between two correlation tensors of three qubit states is defined as
\begin{equation}
(V,W) \equiv \sum_{\mu,\nu,\eta = 0}^3 V_{\mu \nu \eta} W_{\mu \nu \eta}.
\end{equation}

\subsection{Tripartite entanglement}

To keep the statement as general as possible, we prove that all states $\rho^{nc}_{\phi} = \frac{1}{2} \proj{\phi} + \frac{1}{2} \proj{\overline \phi}$ with
\begin{eqnarray}
\ket{\phi} \!& = &\! \sin \beta \cos \alpha \ket{001} + \sin \beta \sin \alpha \ket{010} + \cos \beta \ket{100}, \label{PSI-PSIBAR} \\
\ket{\overline \phi} \!& = &\! \sin \beta \cos \alpha \ket{110} + \sin \beta \sin \alpha \ket{101} + \cos \beta \ket{011}, \nonumber
\end{eqnarray}
are genuinely tripartite entangled as soon as $\ket{\phi}$ is genuinely tripartite entangled. \\
First, note that $\ket{\phi}$ is a bi-product state if at least one amplitude vanishes, i.e., if either
\begin{enumerate}
\item $\beta = 0$ (full product state),
\item $\beta = \frac{\pi}{2}$ and $\alpha = 0$ (full product state),
\item $\beta = \frac{\pi}{2}$ and $\alpha = \frac{\pi}{2}$ (full product state),
\item $\beta = \frac{\pi}{2}$ and $\alpha \in (0,\frac{\pi}{2})$ (bi-product $A|BC$),
\item $\alpha = 0$ and $\beta \in (0,\frac{\pi}{2})$ (bi-product $B|AC$),
\item $\alpha = \frac{\pi}{2}$ and $\beta \in (0,\frac{\pi}{2})$ (bi-product $C|AB$).
\end{enumerate}
The correlation tensor of the state $\rho^{nc}_{\phi}$ contains only bipartite correlations:
\begin{eqnarray}
T_{xx0} & = & T_{yy0} = \sin(2 \beta) \sin(\alpha), \nonumber \\
T_{x0x} & = & T_{y0y} = \sin(2 \beta) \cos(\alpha), \nonumber \\
T_{0xx} & = & T_{0yy} = \sin^2(\beta) \sin(2 \alpha), \nonumber \\
T_{zz0} & = & \cos(2 \alpha) \sin^2(\beta) - \cos^2(\beta), \nonumber \\
T_{z0z} & = & - \cos(2 \alpha) \sin^2(\beta) - \cos^2(\beta), \nonumber \\
T_{0zz} & = & \cos(2 \beta),
\label{T_MIX}
\end{eqnarray}
and $T_{000}=1$. Using these expressions, the right-hand side of the entanglement criterion is
\begin{equation}
R = (T,T) = 4.
\label{eq:R4}
\end{equation}
To find the maximum of the left-hand side, we shall follow a few estimations.
Consider first the bi-product state in a fixed bipartition, say $AB|C$, i.e., of the form $\ket{\chi}_{AB} \otimes \ket{c}$,
where $\ket{\chi}_{AB} = \cos(\theta) \ket{00} + \sin(\theta) \ket{11}$, when written in the Schmidt basis.
Let us denote the correlation tensor of $\ket{\chi}_{AB}$ with $P$ and its local Bloch vectors by $\vec a$ and $\vec b$.
We therefore have:
\begin{eqnarray}
L  &=& 1 + T_{xx0} (P_{xx} + P_{yy}) + T_{zz0} P_{zz} + T_{x0x}(a_x c_x + a_y c_y)  \nonumber \\
&+& T_{z0z} a_z c_z + T_{0xx}(b_x c_x + b_y c_y) + T_{0zz} b_z c_z. 
\label{L}
\end{eqnarray}
By optimizing over the states of $\ket{c}$ we get the following upper bounds:
\begin{equation}
T_{x0x}(a_x c_x + a_y c_y) + T_{z0z} a_z c_z \le \sqrt{T_{x0x}^2 (a_x^2 + a_y^2) + T_{z0z}^2 a_z^2},
\end{equation}
and
\begin{equation}
T_{0xx}(b_x c_x + b_y c_y) + T_{0zz} b_z c_z \le \sqrt{T_{0xx}^2 (b_x^2 + b_y^2) + T_{0zz}^2 b_z^2}.
\label{OVER_C}
\end{equation}
The Schmidt decomposition implies for local Bloch vectors:
\begin{equation}
a_x^2 + a_y^2 + a_z^2 = b_x^2 + b_y^2 + b_z^2 = \cos^2(2 \theta),
\end{equation}
and therefore
\begin{equation}
\vec a = \cos(2 \theta) \vec n, \quad \vec b = \cos(2 \theta) \vec m,
\end{equation}
where $\vec n$ and $\vec m$ are normalized vectors with directions along the local Bloch vectors.
This gives the bound
\begin{eqnarray}
&& \sqrt{T_{x0x}^2 (a_x^2 + a_y^2) + T_{z0z}^2 a_z^2} + \sqrt{T_{0xx}^2 (b_x^2 + b_y^2) + T_{0zz}^2 b_z^2} \nonumber \\
&& =  \cos(2 \theta) ( \sqrt{T_{x0x}^2 (n_x^2 + n_y^2) + T_{z0z}^2 n_z^2} \\
&& +\sqrt{T_{0xx}^2 (m_x^2 + m_y^2) + T_{0zz}^2 m_z^2} ) \nonumber \\
&& \le  \cos(2 \theta) ( \max(|T_{x0x}|,|T_{z0z}|) + \max(|T_{0xx}|,|T_{0zz}|) ),\nonumber
\end{eqnarray}
where the maxima follow from convexity of squared components of a normalized vector.

Now let us focus on the terms depending on the correlations of $\ket{\chi}_{AB}$.
In order to maximize \eqref{L}, the Schmidt basis of $\ket{\chi}_{AB}$ has to be either $x$, $y$, or $z$ as otherwise off-diagonal elements of $P$ emerge leading to smaller values entering \eqref{L}.
For the diagonal correlation tensor we have $|P_{xx}| = \sin(2 \theta)$, $|P_{yy}| = \sin(2 \theta)$, and $P_{zz} = 1$, and with indices permuted.
Therefore, there are three cases to be considered in order to optimize $T_{xx0} (P_{xx} + P_{yy}) + T_{zz0} P_{zz}$:
\begin{itemize}
\item[(i)] $|P_{xx}| = 1$ and $|P_{yy}| = |P_{zz}| = \sin(2 \theta)$ with their signs matching those of $T_{xx0}$ and $T_{zz0}$ respectively,
\item[(ii)] $|P_{zz}| = 1$ and $P_{xx} = P_{yy} = \sin(2 \theta)$,
\item[(iii)] $|P_{zz}| = 1$ and $P_{xx} = - P_{yy} = \sin(2 \theta)$.
\end{itemize}
Each of these cases leads to an upper bound on $L$.
For example, for the first case we find
\begin{widetext}
\begin{eqnarray}
L_{\mathrm{(i)}}  &=&  1 + |T_{xx0}| + \sin(2 \theta) (|T_{xx0}| + |T_{zz0}|)
+ \cos(2 \theta) (\max(|T_{x0x}|,|T_{z0z}|) + \max(|T_{0xx}|,|T_{0zz}|)) \nonumber\\
 &\le&  1 + |T_{xx0}|
+ \sqrt{(|T_{xx0}| + |T_{zz0}|)^2 + (\max(|T_{x0x}|,|T_{z0z}|) + \max(|T_{0xx}|,|T_{0zz}|))^2},
\label{eq:Li}
\end{eqnarray}
\end{widetext}
where in the last step we optimized over $\theta$. The same procedure applied to the other two cases gives:
\begin{widetext}
\begin{eqnarray}
L_{\mathrm{(ii)}} & \le & 1 + |T_{zz0}| + \sqrt{4 T_{xx0}^2 + (\max(|T_{x0x}|,|T_{z0z}|) + \max(|T_{0xx}|,|T_{0zz}|))^2}, \label{eq:Lii}\\
L_{\mathrm{(iii)}} & \le & 1 + |T_{zz0}| + \max(|T_{x0x}|,|T_{z0z}|) + \max(|T_{0xx}|,|T_{0zz}|). \label{eq:Liii}
\end{eqnarray}
\end{widetext}
If instead of the bipartition $AB|C$ another one was chosen, the bounds obtained are given by those above with the indices correspondingly permuted.
Since there are three possible bipartitions, altogether we have nine bounds out of which we should finally choose the maximum as the actual upper bound on the left-hand side.

\subsubsection*{Numerical derivation of bounds}

\begin{figure}[!ht]
\includegraphics[width=0.46\textwidth]{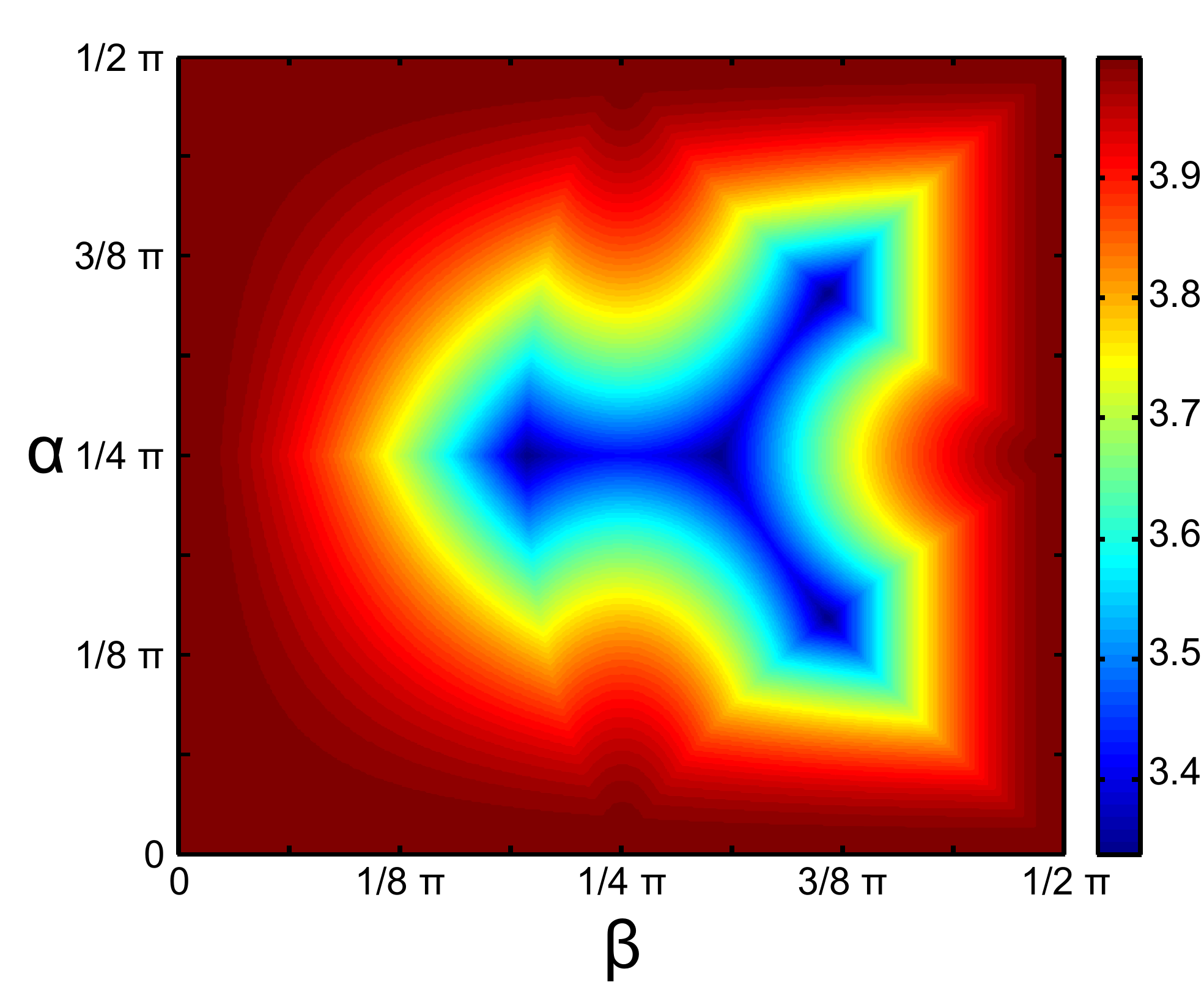}
\caption{Contour plot showing the maximal value of the left-hand side of our entanglement criterion for the states $\rho^{nc}_{\phi}$ defined above (\ref{PSI-PSIBAR}).
Whenever the value is below $4$, i.e., the right-hand side value as given in \eqref{eq:R4}, the criterion detects genuine tripartite entanglement.
This shows that all the states $\rho^{nc}_{\phi}$ are genuinely tripartite entangled except for those arising from bi-product states $\ket{\phi}$, i.e., for $\alpha, \beta = 0$ or $\pi/2$.
Numerical optimizations over all bi-separable states yield the same plot.}
\label{FIG_CRIT}
\end{figure}

A first approach is to numerically evaluate Eqs.~(\ref{eq:Li})-(\ref{eq:Liii}). 
Fig.~\ref{FIG_CRIT} shows that only for states $\ket{\phi}$ that are bi-product the left-hand side reaches $L=4$.

For the $W$ state we thus obtain $\max L = 10/3$ which is achieved by the bi-product state $(\cos\theta \ket{++} - \sin\theta \ket{--}) \otimes \ket{+}$,
where $\ket{\pm} = \frac{1}{\sqrt{2}}(\ket{0} \pm \ket{1})$ and $\tan(2 \theta) = 3/4$ in order to optimize case (i) which is the best for the $W$ state.
This bound is used in the main text.

\subsubsection*{Analytic argument}

The last step of the proof, showing that only bi-separable states can achieve the bound of $4$ in our criterion, involved numerical optimization (Fig.~\ref{FIG_CRIT}). 
One may complain that due to finite numerical precision there might be genuinely tripartite entangled states for values of $\alpha$ or $\beta$ close to $0$ and $\pi/2$ that already achieve the bound of $4$.
Here, we give a simple analytical argument showing that $\rho^{nc}_{{\phi}}$ is genuinely tripartite entangled if and only if $\ket{\phi}$ is so.

We first follow the idea of Ref.~\cite{PhysRevLett.101.070502} and note that a mixed state  $\rho^{nc}_{\phi}$ can only be bi-separable if there are bi-product pure states in its support.
The support of $\rho^{nc}_{\phi}$ is spanned by $\ket{\phi}$ and $| \overline{\phi} \rangle$, i.e., $\rho^{nc}_{\phi}$ does not have any overlap with the orthogonal subspace $\openone - \proj{\phi} - | \overline{\phi} \rangle \langle \overline{\phi} |$. 
Accordingly any decomposition of $\rho^{nc}_{\phi}$ into pure states can only use pure states of the form
\begin{equation}
\ket{\Phi} = a \ket{\phi} + b | \overline{\phi} \rangle.
\end{equation}
We now give a simple argument that $\ket{\Phi}$ is bi-product, and hence $\rho^{nc}_{\phi}$ is bi-separable, if and only if $|\phi\rangle$ is bi-product.
In all other infinitely many cases, the no-correlation state is genuinely tripartite entangled.
Assume that $|\Phi\rangle$ is bi-product in the partition $AB|C$. 
Accordingly, all its correlation tensor components factor across this partition.
In particular, 
\begin{eqnarray}
&&T_{0xx} = W_{0x} V_x, \,\,\,\,\,\,\, T_{0yy} = W_{0y} V_y, \,\,\,\,\,\,\, \\
&&T_{0xy} = W_{0x} V_y, \,\,\,\,\,\,\,  T_{0yx} = W_{0y} V_x \nonumber
\end{eqnarray}
where $W$ is the correlation tensor of the state of $AB$ and $V$ is the correlation tensor corresponding to the state of $C$.
One directly verifies that for such a bi-product state we have 
\begin{equation}
T_{0xx} T_{0yy} = T_{0xy} T_{0yx}.
\label{T_BIPROD}
\end{equation}
Evaluating condition (\ref{T_BIPROD}) for the states $\ket{\Phi}$ gives the following condition on the amplitudes of $|\phi\rangle$:
\begin{equation}
\sin^2(2 \alpha) \sin^4(\beta) = 0,
\end{equation}
and indicates that at least one amplitude must be zero.
Similar reasoning applies to other partitions and we conclude that $\ket{\Phi}$ is bi-product if and only if $\ket{\phi}$ is bi-product.

\subsubsection*{Alternative entanglement criterion}

Alternativly we can apply a witness of genuine tripartite entanglement based on angular momentum operators
~\cite{JOptSocAmB.24.275},
\begin{equation}
\mathcal{W}_3 = J_x^2 + J_y^2,
\end{equation}
where e.g. $J_x = \frac{1}{2}(\sigma_x \otimes \openone \otimes \openone + \openone \otimes \sigma_x \otimes \openone + \openone \otimes \openone \otimes \sigma_x)$.
Maximization of this quantity over bi-separable states gives~\cite{JOptSocAmB.24.275}:
\begin{equation}
\max_{\rho^{\mathrm{bi-sep}}} \langle \mathcal{W}_3 \rangle = 2+\sqrt{5}/2 \approx 3.12.
\label{J_BISEP}
\end{equation} 
This criterion detects entanglement of the states $\ket{\phi}$ and $\ket{\overline \phi}$, and, consequently, 
since it uses two-party correlations only, also of the state $\rho_{\phi}^{nc}$. However, entanglement
is detected only for a range of roughly $\alpha \in [0.59,1.3]$ and $\beta \in [0.33,1.2]$.

\subsection{Five-partite entanglement}

In order to obtain the five-partite bound given in the main text, i.e., $\max_{T^{{bi-prod}}} (T,T^{{bi-prod}}) = 12.8$,
we have numerically optimized over all bi-product states keeping $T$ as the correlation tensor of an equal mixture of Dicke states $| D_5^{(2)} \rangle$ and $|D_5^{(3)} \rangle$, where
\begin{equation}
| D_n^{(e)} \rangle = \frac{1}{\sqrt{{n \choose e}}} \sum_i | \mathcal{P}_i(1,\dots,1,0\dots,0) \rangle,
\end{equation}
with $\mathcal{P}_i$ denoting all distinct permutations of $e$ ones and $n-e$ zeros.

Below, we generalize the analytical argument given above to prove genuine multipartite entanglement of arbitrary mixtures of Dicke and anti-Dicke states.
The anti-Dicke state has exchanged roles of zeros and ones as compared with the Dicke state, i.e., it has $n-e$ ones (excitations).
One easily verifies that the Dicke state of $n$ qubits with $e$ excitations has the following bipartite correlations:
\begin{eqnarray}
T_{0 \dots 0 xx} & = & T_{0 \dots 0 yy} = \frac{2 {n-2 \choose e-1}}{{n \choose e}} = \frac{2 e (n-e)}{n (n-1)}, \nonumber \\
T_{0 \dots 0 xy} & = & T_{0 \dots 0 yx} = 0.\label{D-DBAR}
\end{eqnarray}
The correlations of an anti-Dicke state, with $n-e$ excitations, are the same due to the symmetry $e \leftrightarrow n-e$ of these correlations.
Assume that $n$ is odd so that (i) the Dicke and anti-Dicke states are orthogonal and (ii) the parity of the number of excitations, i.e., whether there is an even or odd number of them, is opposite in the Dicke and anti-Dicke states.
For arbitrary superposition $\alpha | D_n^{(e)} \rangle + \beta | D_n^{(n-e)} \rangle$ the correlations read:
\begin{eqnarray}
T_{0 \dots 0 jk} &=& |\alpha|^2 T_{0 \dots 0 jk}^D + |\beta|^2 T_{0 \dots 0 jk}^{\overline{D}} \nonumber \\
 &+& \alpha^* \beta \langle D_n^{(e)} | \openone \otimes \dots \openone \otimes \sigma_j \otimes \sigma_k | D_n^{(n-e)} \rangle \\
&+& \alpha \beta^* \langle D_n^{(n-e)} | \openone \otimes \dots \openone \otimes \sigma_j \otimes \sigma_k | D_n^{(e)} \rangle. \nonumber
\end{eqnarray}
Since applying $\sigma_j \otimes \sigma_k$ with $j,k=x,y$ to the Dicke states does not change the parity of their excitations, the last two terms vanish, and for the first two terms we have $T_{0 \dots 0 jk}^D = T_{0 \dots 0 jk}^{\overline{D}}$.
Therefore, an arbitrary superposition of Dicke and anti-Dicke states has the same correlations as in (\ref{D-DBAR}) and therefore none of such superposed states is bi-product.
Since the Dicke states are invariant under exchange of parties (and so are their superpositions), the same holds for other partitions.
Finally, the lack of bi-product states in a subspace spanned by Dicke and anti-Dicke states implies that their mixtures are also genuinely multipartite entangled.

\section{Genuine tripartite correlations}

\label{C}

\begin{figure}[!ht]
\includegraphics[width=0.49\textwidth]{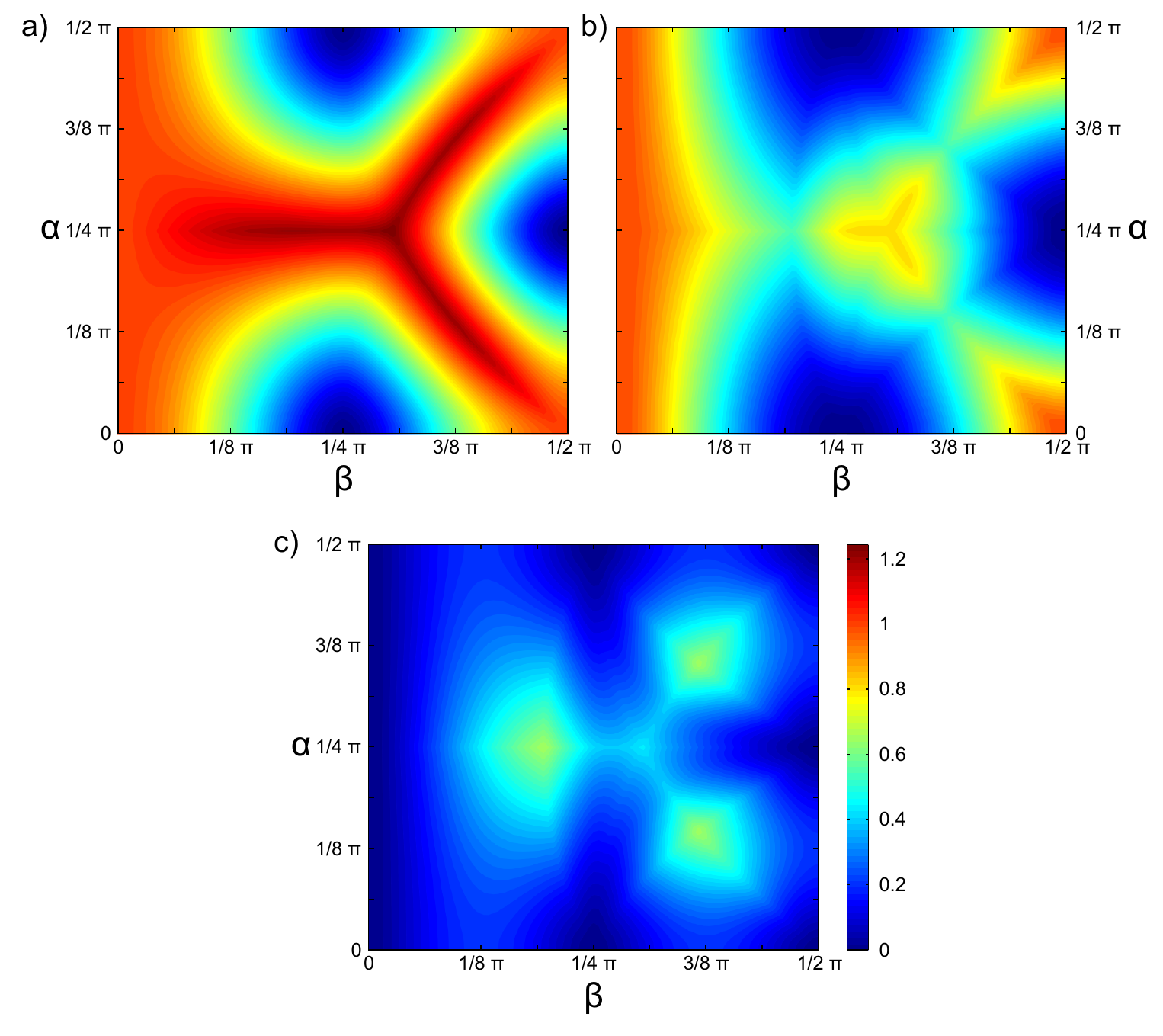}
\caption{\label{ecor}
Correlation content~\cite{PhysRevLett.107.190501} of the states $\rho^{nc}_{\phi} = \frac{1}{2} \proj{\phi} + \frac{1}{2} \proj{\overline \phi}$ with the pure states given in Eq.~(\ref{PSI-PSIBAR}).
(a) \emph{Total genuine tripartite correlations}.
The genuine tripartite correlations vanish only for mixtures of bi-product states. The highest value ($1.2516$) is obtained for the state $(|W\rangle \langle W |+  |\overline{W}\rangle \langle \overline{W}|)/2$.
(b) {\em Genuine tripartite classical correlations}. The genuine classical correlations also vanish only for mixtures of bi-product states. The highest value (1.0) is observed for fully separable states. The local maximum (0.8127) is achieved by the state $(|W\rangle \langle W |+  |\overline{W}\rangle \langle \overline{W}|)/2$.
(c) {\em Genuine tripartite quantum correlations}. The genuine quantum correlations vanish for mixtures of bi-product states and for fully separable states.  The highest values (0.6631) correspond to the mixture of the state
$\sqrt{1/6} |001\rangle + \sqrt{1/6}|010\rangle + \sqrt{2/3}|100\rangle$ with its antistate (and permutations). The state $(|W\rangle \langle W |+  |\overline{W}\rangle \langle \overline{W}|)/2$ achieves the local maximum (0.4389).
}
\end{figure}

While the conventional full correlation function vanishes for $\rho^{nc}_{\phi}$, this is not necessarily so for other types of correlation functions introduced recently.
For a comparison we analyze the correlation content of the states of our family also according to the three measures given in Ref. \cite{PhysRevLett.107.190501},
namely: (a)  genuine tripartite correlations $T^{(3)}(\rho^{nc}_{\phi})$,
(b) genuine tripartite classical correlations $J^{(3)}(\rho^{nc}_{\phi})$,
and (c) genuine tripartite quantum correlations $D^{(3)}(\rho^{nc}_{\phi})$.
The results are presented and discussed in Fig. \ref{ecor}.

\section{Experimental three and five qubit states}

\label{D}

\begin{figure*}[!ht]
\includegraphics[width=0.8\textwidth]{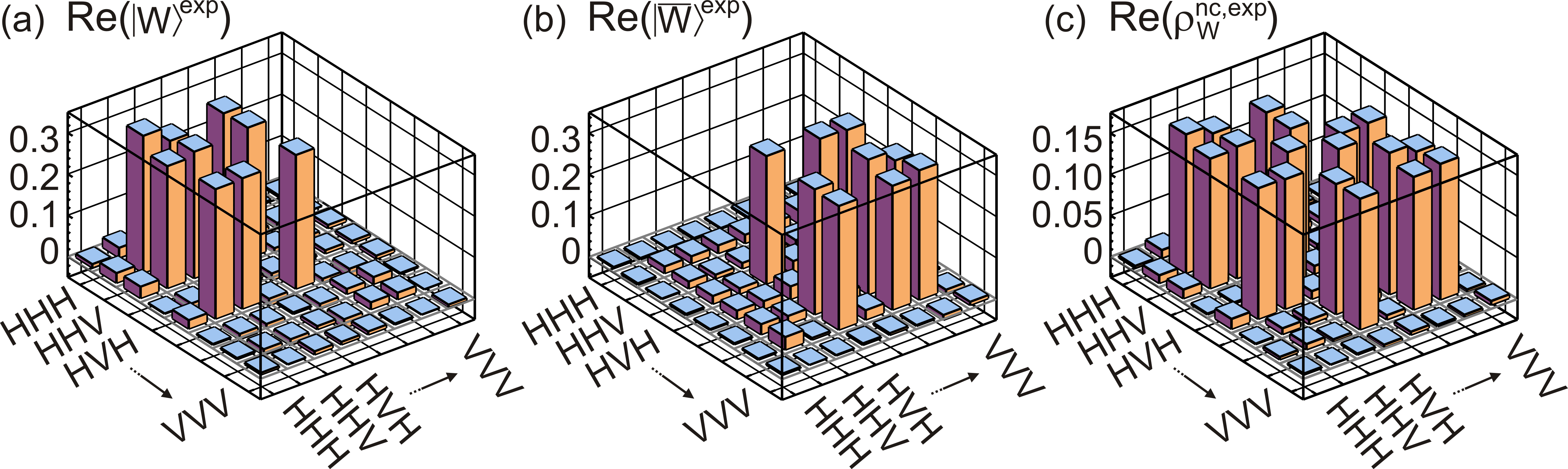}
\caption{\label{3QUBITS}
Experimental three qubit states as obtained from the state $|D_4^{(2)}\rangle^{exp}$. (a) The state $\ket{W}^{exp}$ is obtained by projection of the fourth qubit of $|D_4^{(2)}\rangle^{exp}$ on $V$. (b)
The state $\ket{\overline{W}}^{exp}$ is prepared by projecting the fourth qubit of $|D_4^{(2)}\rangle^{exp}$ on $H$. (c) When the fourth qubit of $|D_4^{(2)}\rangle^{exp}$ is traced out, a mixture of
$\ket{W}^{exp}$ and $\ket{\overline{W}}^{exp}$ is obtained, i.e., the state $\rho_W^{nc,exp}$.
The corresponding fidelities with respect to their target states are  $0.939\pm0.011$ for $\ket{W}^{exp}$, $0.919\pm0.010$ for $\ket{\overline{W}}^{exp}$, and $0.961\pm0.003$ for $\rho^{nc,exp}_W$.
}
\end{figure*}

\begin{figure}[!ht]
\includegraphics[width=0.48\textwidth]{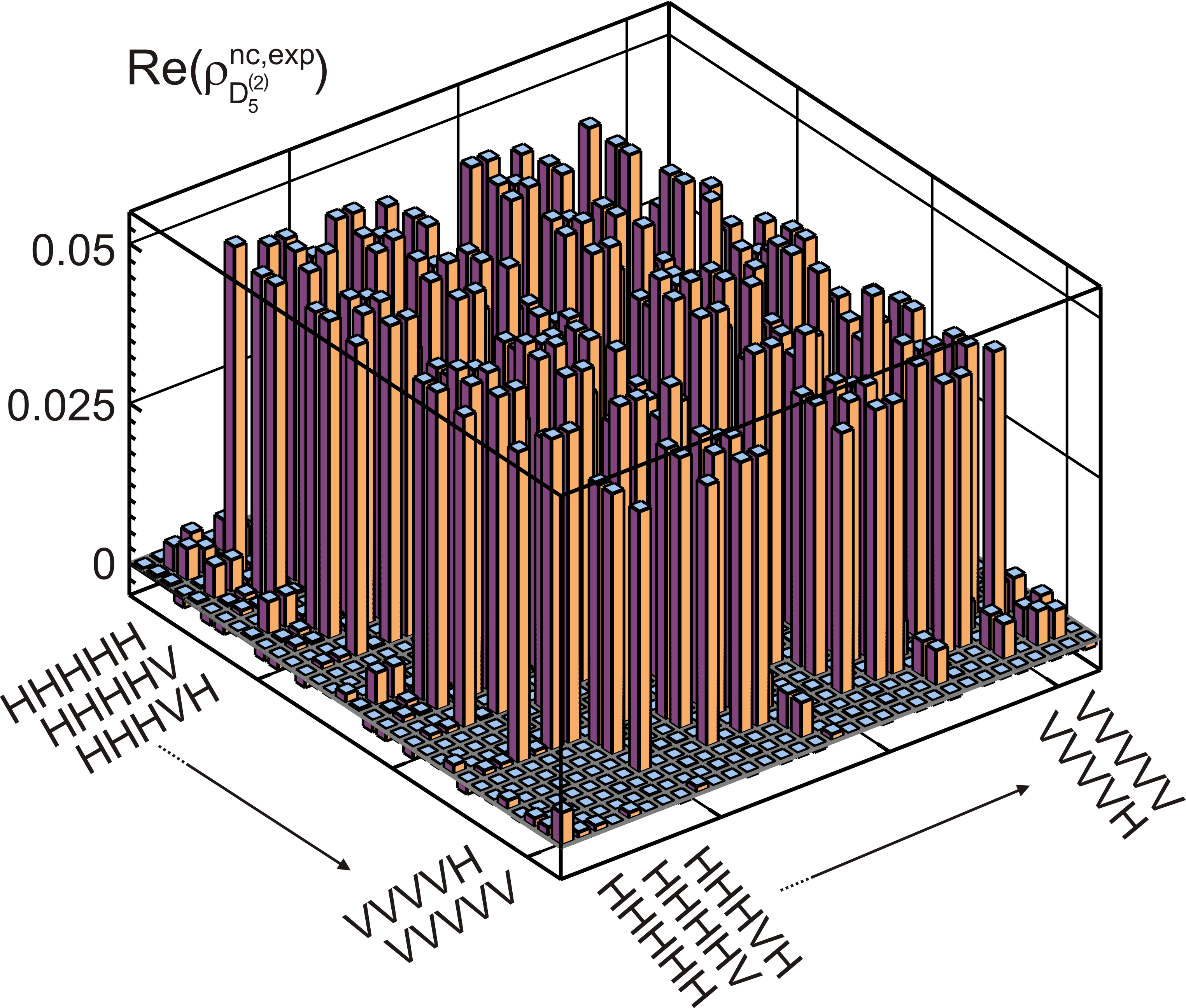}
\caption{\label{5QUBITS}
Experimental state $\rho^{nc,exp}_{D_{5}^{(2)}}$ determined from five-fold coincidences together with permutational invariant tomography~\cite{PhysRevLett.113.040503}.
The fidelity with respect to the target state is $0.911\pm0.004$.
}
\end{figure}

The experimentally prepared states $\ket{W}^{exp}$, $\ket{\overline{W}}^{exp}$, $\rho_W^{nc,exp}$, and $\rho^{nc,exp}_{D_{5}^{(2)}}$ were characterized by means of quantum state tomography. Their corresponding density matrices can be seen in Fig.~\ref{3QUBITS} and Fig.~\ref{5QUBITS}.
The fidelities of the observed three qubit states with respect to their target states are  $0.939\pm0.011$ for $\ket{W}^{exp}$, $0.919\pm0.010$ for $\ket{\overline{W}}^{exp}$, and $0.961\pm0.003$ for $\rho^{nc,exp}_W$. 
Note that the value of the fidelity for the state $\rho^{nc,exp}_W$ was obtained from a maximum likelihood (ML) reconstruction together with non-parametric bootstrapping. This value thus might be slightly incorrect due to the bias of the maximum likelihood data evaluation~\cite{arxiv}.

Fig.~\ref{5QUBITS} shows the real part of the tomographically determined no-correlation state from which all further five qubit results are deduced.
The five-qubit fidelity of $\rho^{nc,exp}_{D_{5}^{(2)}}$ is determined via a ML reconstruction from five-fold coincidences to be $0.911\pm0.004$.

\begin{figure}[!ht]
\includegraphics[width=0.46\textwidth]{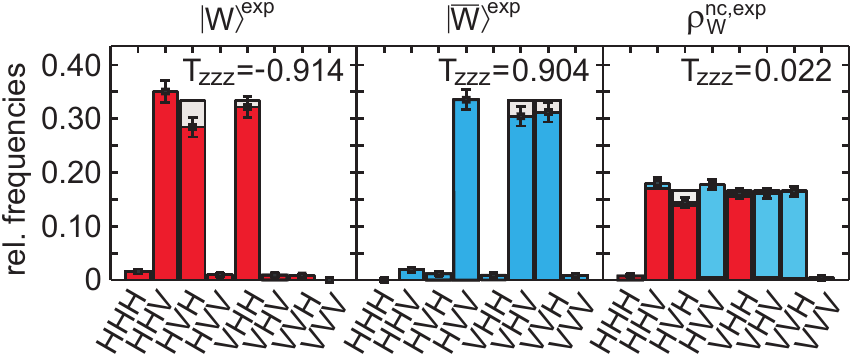}
\caption{(color online). Detection frequencies when observing the states $\ket{W}^{exp}$ (red) and $\ket{\overline{W}}^{exp}$ (blue) and $\rho_W^{nc,exp}$ (red and blue) in the $\sigma_z^{\otimes 3}$ basis. From these data $T_{zzz}$ values can be calculated showing how the correlations of $\ket{W}^{exp}$ and $\ket{\overline{W}}^{exp}$ average to approximately 0. For comparison, the theoretically expected values are shown in gray. 
The correlation value $T_{zzz}$ of the state $\rho_W^{nc,exp}$ was determined as the weighted sum of the correlation values $T_{zzz}$ of the states $\ket{W}^{exp}$ and $\ket{\overline{W}}^{exp}$. The state $\ket{W}^{exp}$ was observed with a slightly lower probability ($0.485$) than the state $\ket{\overline{W}}^{exp}$ ($0.515$) leading to a value of $T_{zzz} = 0.022$ for the state $\rho_W^{nc,exp}$.
In contrast, in Fig.~2 of the main text the states $\ket{W}^{exp}$ and $\ket{\overline{W}}^{exp}$ were obtained from the state $| D_{4}^{(2)} \rangle^{exp}$ by projection of the fourth qubit onto horizontal/vertical polarization, i.e., from measuring $\sigma_z$ on the fourth qubit. There, $\rho_W^{nc,exp}$ was obtained by tracing out the fourth qubit and hence measurements of $\sigma_x, \sigma_y, \sigma_z$ on the fourth qubit of $| D_{4}^{(2)} \rangle^{exp}$ contribute, leading to approximately three times better statistics for the state $\rho_W^{nc,exp}$.    
}
\label{FIG_CORRS}
\end{figure}

To obtain a correlation function value, e.g., $T_{zzz}=\operatorname{Tr}(\rho~ \sigma_z \otimes \sigma_z \otimes \sigma_z)$, we analyze the three photons in the respective set of bases (here all $\hat{z}$). Fig.~\ref{FIG_CORRS} shows the relative frequencies for observing all the possible results for such a polarization analysis. Clearly one recognizes the complementary structure of the the detection frequencies for the states $\ket{W}^{exp}$ and $\ket{\overline{W}}^{exp}$ which results in approximately the same magnitude of the correlations, yet with different sign. Mixing the two states, one thus obtains a vanishingly small correlation. Fig.~[2] of the main text then shows the full set of correlations.

\begin{figure}[!ht]
\includegraphics[width=0.49\textwidth]{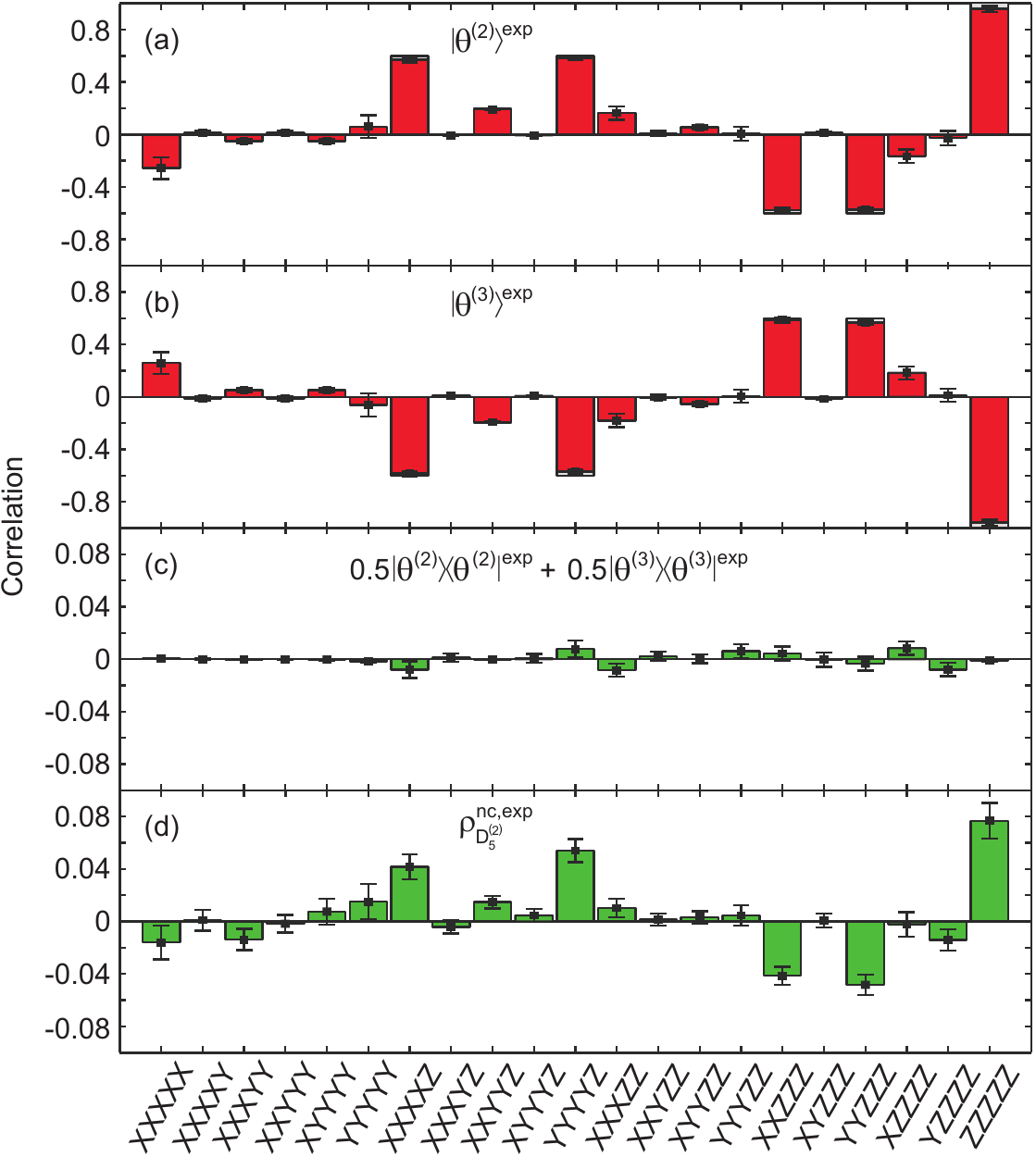}
\caption{\label{FIG_5P}
Experimental five-partite symmetric correlations for the two most prominent states (a) $|\Theta^{(2)} \rangle^{exp}$ and (b) $|\Theta^{(3)} \rangle^{exp}$ in the eigen decomposition of the experimental density matrix $\rho^{nc,exp}_{D_{5}^{(2)}}$ shown in Fig. \ref{5QUBITS}. The correlations of these states are compared with the ones of the states (a) $| D_{5}^{(2)} \rangle$ and (b) $| D_{5}^{(3)} \rangle$, respectively, shown in gray. The agreement between the actual and expected correlations is evident and also the fidelities of $|\Theta^{(2)} \rangle^{exp}$ and  $|\Theta^{(3)} \rangle^{exp}$ with the respective target states are high: $F_{| D_{5}^{(2)} \rangle}(| \Theta^{(2)} \rangle^{exp})=0.978\pm0.012$ and $F_{| D_{5}^{(3)} \rangle}(| \Theta^{(3)} \rangle^{exp})=0.979\pm0.012$. (c) When both states are evenly mixed, the resultant state has practically vanishing correlations. (d) Since the collection efficiencies for signal and idler photons generated via spontaneous parametric down-conversion differ 
slightly \cite{SignalIdler}, the 
states  $|\Theta^{(2)} \rangle^{exp}$ and $|\Theta^{(3)} 
\rangle^{exp}$ are observed with relative weights of $0.54$ and $0.46$ leading to largely suppressed but not entirely vanishing full correlations. Hence, the experimentally prepared state $\rho_{D_5^{(2)}}^{{nc,exp}}$ is a very good approximation to a no-correlation state. Please note that the correlations shown in (c) and (d) are magnified by a factor of $10$ compared with the scale of (a) and (b).
The errors given in subfigures (a)-(c) were obtained by non-parametric bootstrapping~\cite{bootstrapping} whereas for (d) Gaussian error propagation was used.
}
\end{figure}

For the analysis of the five qubit no correlation state, we see from an eigen decomposition that this state indeed comprises of a mixture of two states ($|\Theta^{(2)} \rangle^{exp}$ and $|\Theta^{(3)} \rangle^{exp}$), which are in very good agreement with  $| D_{5}^{(2)} \rangle$ and $| D_{5}^{(3)} \rangle$.
Fig.~\ref{FIG_5P} (a) and (b) show all symmetrized correlations for the five-qubit states $| \Theta^{(2)} \rangle$ and $| \Theta^{(3)} \rangle$ and $\rho_{D_5^{(2)}}^{{nc,exp}}$ with good agreement with the ideal states. 
Also the respective fidelity of the eigenvectors of the experimentally determined state are quite high ($F_{| D_{5}^{(2)} \rangle}(| \Theta^{(2)} \rangle)=0.978\pm0.012$ and $F_{| D_{5}^{(3)} \rangle}(| \Theta^{(3)} \rangle^{exp})=0.979\pm0.012$). Equally mixing the states $|\Theta^{(2)} \rangle^{exp}$ and $|\Theta^{(3)}\rangle^{exp}$ indeed would result in a state with vanishingly small correlations as seen in  Fig.~\ref{FIG_5P} (c). However, due to asymmetry in the coupling of signal and idler states from the down conversion source~\cite{SignalIdler} the correlations are still present, albeit smaller by a factor of 10 compared with $| D_{5}
^{(2)} \rangle$ and $| D_{5}^{(3)} \rangle$.
In the main text we show that the very same state is genuinely five-party entangled.


\section{Statistical analysis}
\label{E}

\subsection{Error analysis}

In order to carry out $n$-qubit quantum state tomography, we measured in the eigenbases of all $3^n$ combinations of local Pauli settings $s_i$ with $s_1 = x...xx$,
$s_2 = x...xy$, ..., $s_{3^n} = z...zz$.
In each setting $s_i$ we performed projection measurements on all the $2^n$ eigenvectors of the corresponding operators.
The single measurement results are enumerated by $r_j$ representing the binary numbers from $0$ to $2^n-1$ in increasing order, i.e., $r_1 = 0...00$, $r_2 = 0...01$, ..., $r_{2^n} = 1...11$.
The observed counts for the outcome $r_j$ when measuring $s_i$ are labeled as $c_{r_j}^{s_i}$ and the total number of counts $N_{s_i}$ for setting $s_i$ is given by $N_{s_i} = \sum\limits_{j=1}^{2^n} c_{r_j}^{s_i}$.
From these data the density matrix can be obtained as
\begin{equation}
\rho = \sum\limits_{i=1}^{3^n}\sum\limits_{j=1}^{2^n} \frac{c_{r_j}^{s_i}}{N_{s_i}} M_{r_j}^{s_i}
\end{equation}
where the elements of the generating set of operators $M_{r_j}^{s_i}$ are defined as $M_{r_j}^{s_i} = \frac{1}{2^n}\bigotimes\limits_{k=1}^n \Big(\frac{\identity}{3}+(-1)^{r_j(k)}\sigma_{s_i(k)}\Big)$~\cite{James,PhDNikolai}, where $\identity$ denotes the $2\times2$ identity matrix and $r_{j(k)}$ is the k-th entry in the string $r_j$.
Then, the fidelity $F_{\ket{\psi}}$ with respect to a pure target state $\ket{\psi}$ can be calculated as
\begin{equation}
F_{\ket{\psi}} = \bra{\psi}\rho\ket{\psi} = \sum\limits_{i=1}^{3^n}\sum\limits_{j=1}^{2^n}  \frac{c_{r_j}^{s_i}}{N_{s_i}} \bra{\psi} M_{r_j}^{s_i} \ket{\psi}.
\end{equation}
For Poissonian measurement statistics, i.e., $\Delta c_{r_j}^{s_i} = \sqrt{c_{r_j}^{s_i}}$, the error to the fidelity $\Delta F_{\ket{\psi}} = \sqrt{\Delta^2 F_{\ket{\psi}}}$ can be deduced via Gaussian error propagation as $\Delta^2 F_{\ket{\psi}} = \sum\limits_{i=1}^{3^n}\sum\limits_{j=1}^{2^n}  (\frac{1}{N_{s_i}} - \frac{1}{N_{s_i}^2})^2\bra{\psi} M_{r_j}^{s_i} \ket{\psi}^2 c_{r_j}^{s_i}$
which is approximately
\begin{equation}
\Delta^2 F_{\ket{\psi}} = \sum\limits_{i=1}^{3^n}\Delta^2 F_{\ket{\psi}}^{s_i} = \sum\limits_{i=1}^{3^n}\sum\limits_{j=1}^{2^n}  \frac{c_{r_j}^{s_i}}{N_{s_i}^2}\bra{\psi} M_{r_j}^{s_i} \ket{\psi}^2
\label{eq:approxerr}
\end{equation}
for large number of counts per setting as in our experiment.
As an example, in table \ref{tab:zzz} we give the corresponding values for $c_{r_j}^{s_i}$ and $|\langle \psi | M_{r_j}^{s_i}| \psi \rangle |$ for the $2^3=8$ possible results of the $zzz$ measurement of the three qubit $\ket{W}$ state to get an impression of the size of the $3^3=27$ terms in Eq.~(\ref{eq:approxerr}).\\

\renewcommand{\arraystretch}{1.4}

\begin{table*}[!ht]
	  \begin{tabular*}{129mm}{l|c|r|r|r|r|r|r|r|r}
    \hline\hline
      & $r_j$ & $000$ & $001$ & $001$ & $011$ & $100$ & $101$ & $110$ & $111$ \\\cline{1-10}
      $zzz$& $ |\langle \psi | M_{r_j}^{zzz} | \psi \rangle |$ & 1.48e-01 & 1.48e-01 & 1.48e-01 & 1.11e-01 & 1.48e-01 & 1.11e-01 & 1.11e-01 & 7.41e-02 \\\cline{2-10}
      & counts $c_{r_j}^{zzz}$ & 14 & 309 & 250 & 8.71 & 283 & 8 & 7.07 & 0 \\\cline{2-10}
    \hline\hline			
  \end{tabular*}
\caption{\label{tab:zzz} The values of $c_{r_j}^{s_i}$ and $|\langle \psi | M_{r_j}^{s_i}| \psi \rangle |$ for the measurement of the setting $zzz$ of the experimentally observed state $\ket{W}^{{exp}}$.
  The first row shows all possible results $r_j$ associated with the eigenvectors on which projection measurements are performed, labeled in binary representation.
  Please note that the observed counts $c_{r_j}^{s_i}$ are not integers since the slightly differing relative detection efficiencies of the single photon counters were included.
	From these data we obtain for $s_i=zzz$ a contribution for Eq.~(\ref{eq:approxerr}) of $\Delta^2 F_{\ket{W}}^{zzz} = 2.46$e-05.}
   \end{table*}

Similarly, also the error of the $4^3=64$ correlations of the given state are evaluated. For example, we obtain for the correlation value $T_{zzz} = -0.914 \pm 0.034$.
The error for the maximum likelihood estimate was determined by non-parametric bootstrapping, for details see~\cite{bootstrapping}.

\subsection{Hypothesis testing}
\label{sec:hypotheses}
\subsubsection*{Vanishing correlations}
After having calculated the experimental error of the $zzz$ correlation, we find that the measurements of the remaining 26 full correlations have similar errors.
We test our hypothesis of vanishing full correlations by comparing our measured correlation values with a normal distribution with mean $\mu=0$ and standard deviation $\sigma=0.0135$, which corresponds to the average experimental standard deviation.
If our data are in agreement with this distribution, we can retain the hypothesis of vanishing full correlations. \\
\begin{figure}[!ht]
\includegraphics[width=0.49\textwidth]{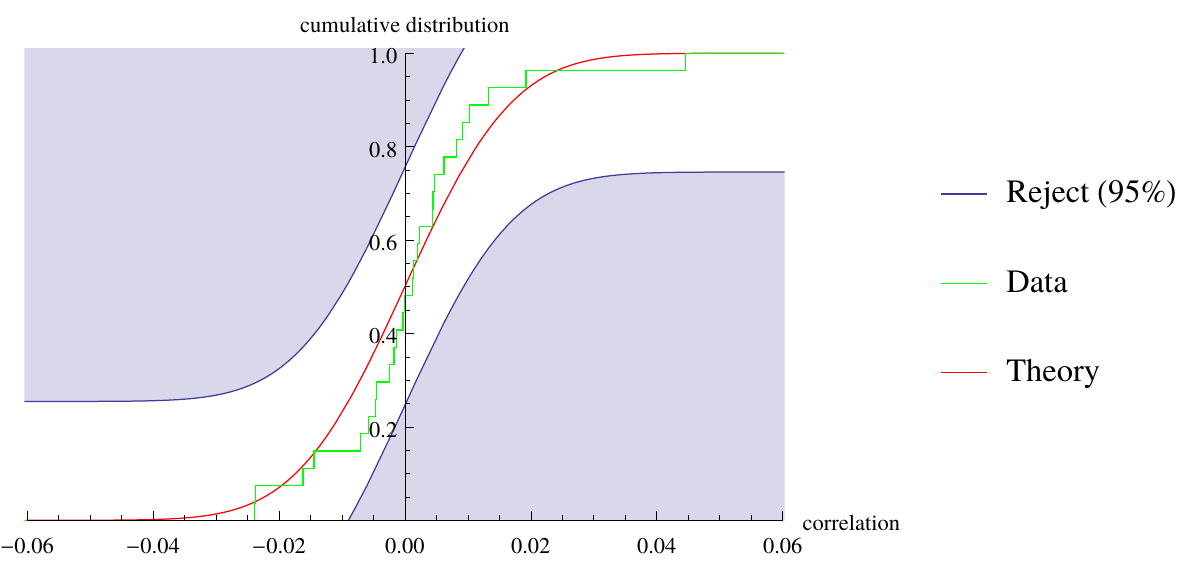}
\caption{The cumulative distribution of the experimentally determined correlations is compared to the cumulative distribution of the expected correlations ($\mu=0$, $\sigma=0.0135$).
The shaded blue region contains points that would be sampled from the normal distribution with probability smaller than $5\%$.
Since the empirical function lies in between the shaded regions, our hypothesis of vanishing correlations can be retained with significance level of $0.05$.}
\label{HypothesisKS}
\end{figure}

To test the hypothesis
\begin{quote}
$H_0^{(nc,3)}:$ all full correlations of the state $\rho_W^{{nc,exp}}$ vanish,
\end{quote}
according to the Kolmogorov-Smirnov method, the cumulative distribution of the $27$ measured full correlations is compared with the cumulative probability distribution of the assumed normal distribution, see Fig.~\ref{HypothesisKS}, quantifying the hypothesis of vanishing full correlations.
We can directly see that the data do not enter the region of rejection given by a significance level of $0.05$.
This clearly indicates that the hypothesis of normal distribution with mean $\mu=0$ and $\sigma=0.135$ cannot be rejected.
While this test (Kolmogorov-Smirnov hypothesis test) is demonstrative, the Anderson-Darling test is considered to be more powerful, i.e., to decrease the probability of errors of second kind.
Since the Anderson-Darling test gives a $p$-value of $0.44$ far above a $0.05$ significance level, we can retain the claim that our measured data indeed correspond to vanishing full correlations, while their scatter can be fully explained by the experimental error. 
%
%

\subsubsection*{Testing for genuine multipartite entanglement}
Furthermore, we also check our hypotheses of the main text that the tripartite and five-partite states are genuinely multipartite entangled.
For that purpose, we calculate the probability that a state without genuine multipartite entanglement achieves values comparable to the measured value based on the assumption that the measurement errors are normally distributed.
Let us formulate for the tripartite state the null hypothesis
\begin{quote}
$H_0^{(3)}:$ state $\rho_W^{{nc,exp}}$ is not genuinely tripartite entangled.
\end{quote}
To show the genuine tripartite entanglement of that state, we want to reject the null hypothesis $H_0^{(3)}$.
In order to estimate the error of first kind, i.e., the probability that $H_0^{(3)}$ is \textit{true}, we calculate the probability that a state without tripartite entanglement achieves the measured value of $\left(T,T_W^{{nc,exp}}\right)=3.858$.
The calculation is based on the assumption of a normal distributed result of the indicator with mean $\mu=\frac{10}{3}$, i.e., the bi-separable bound, and with standard deviation given by our experimental error of $\sigma=0.079$.
The probability of the error of first kind is then at most
\begin{eqnarray}
p&=&\operatorname{Pr}\left[\left(T,T_W^{{nc,exp}}\right)\geq3.858\Big|H_0^{(3)}\right] \\
&<&\frac{1}{\sqrt{2\pi}\sigma} \int_{3.858}^{\infty} {\rm d}x \exp\left({-\frac{\left(x-\mu\right)^2}{2\sigma^2}}\right) \nonumber \\
&=& 1.55\times10^{-11} \ll 0.05. \nonumber
\end{eqnarray}
Since $p$ is far below the significance level of $0.05$, our experimentally implemented state $\rho^{nc}_{W}$ is genuine tripartite entangled. \\

Analogously, we test if the state $\rho_{D_5^{(2)}}^{{nc,exp}}$ is indeed genuinely five-partite entangled.
For that purpose, we formulate the null hypothesis
\begin{quote}
$H_0^{(5)}:$ state $\rho_{D_5^{(2)}}^{{nc,exp}}$ is not genuinely five-partite entangled.
\end{quote}
In order to test the probability that a bi-separable state can achieve $\left(T,T_{D_5^{(2)}}^{{nc,exp}}\right)=13.663$, we now use a normal distribution centered around the bi-separable bound of $\mu=12.8$.
The standard deviation is chosen according to the experimental error of $\sigma=0.340$, such that the probability for a false rejection of the null hypothesis $H_0^{(5)}$ is estimated to be at most
\begin{eqnarray}
p&=&\operatorname{Pr}\left[\left(T,T_{D_5^{(2)}}^{{nc,exp}}\right)\geq13.663\Big|H_0^{(5)}\right] \\
&<&\frac{1}{\sqrt{2\pi}\sigma} \int_{13.663}^{\infty} {\rm d}x \exp\left({-\frac{\left(x-\mu\right)^2}{2\sigma^2}}\right) \nonumber\\
&=&5.6\times10^{-3} \ll 0.05, \nonumber
\end{eqnarray}
clearly indicating the five-partite entanglement of our state with high significance.

\subsubsection*{Bell inequality}
Finally, we test whether we can retain our claim that the five-partite state is non-classical due to its violation of the Bell inequality.
In order to show the violation, we formulate the null hypothesis
\begin{quote}
$H_0^{B}:$ violation of the Bell inequality can be explained by LHV model (finite statistics loophole).
\end{quote}
For the considered Bell inequality~\cite{PhysRevA.86.032105}
\begin{eqnarray}
{\cal B}&=&E_{\mathcal{P}\left(11110\right)}+E_{\mathcal{P}\left(22220\right)}+E_{\mathcal{P}\left(12220\right)} \\
&-&E_{\mathcal{P}\left(21110\right)}-E_{\mathcal{P}\left(11000\right)}-E_{\mathcal{P}\left(22000\right)}\leq6 \nonumber
\end{eqnarray}
with $\mathcal{P}$ denoting the summation over all permutations, e.g. $E_{\mathcal{P}\left(11110\right)}=E_{11110}+E_{11101}+E_{11011}+E_{10111}+E_{01111}$, we calculate the probability that an LHV model can achieve the measured value of ${\cal B}=6.358$, which was estimated with a standard deviation of $\Delta {\cal B}=0.149$.
Following Ref.~\cite{arxiv14070363} we assume that the LHV model gives the maximal allowed expectation value of our Bell parameter, equal to $\mu = 6$, and that the standard deviation of a normal distribution about this mean value is equal to our experimental standard deviation $\Delta {\cal B}$.
Therefore, the probability that the LHV model gives values at least as high as observed is found to be
%
%
%
\begin{eqnarray}
p&=&\operatorname{Pr}\left[{\cal B}\geq6.358\Big|H_0^{B}\right] \\
&<& \frac{1}{\sqrt{2\pi}\sigma} \int_{6.358}^{\infty} {\rm d}x \exp\left({-\frac{\left(x-\mu\right)^2}{2\sigma^2}}\right) = 0.0083 \ll 0.05. \nonumber
\end{eqnarray}
This small $p$-value clearly indicates that the null hypothesis $H_0^{B}$ is to be rejected and thus the non-classicality of the no-correlation state is confirmed.


\subsection{Vanishing full correlations with arbitrary measurement directions}
The measurements presented in the main text show not only vanishing full correlations for measurements in $x$, $y$, $z$ directions, but also for measurements of one qubit rotated in the $yz$-plane. 
Here, we show that full correlations have to vanish for arbitrary measurement directions.
Since the $2$-norm of the correlation tensor is invariant under local rotations, its entries vanish in all local coordinate systems if they do in one. 
Moreover, $l$-fold correlations in one set of local coordinate system only depend on $l$-fold correlations of another set.
As an example, we explicitly show this for the case of three qubits.
\begin{equation}
T_{(\theta_1,\phi_1)\,(\theta_2,\phi_2)\,(\theta_3,\phi_3)}={\rm Tr}(\rho ~\sigma_{(\theta_1,\phi_1)} \otimes \sigma_{(\theta_2,\phi_2)} \otimes  \sigma_{(\theta_3,\phi_3)}) 
\end{equation}
with
\begin{equation}
\sigma_{(\theta_i,\phi_i)} = \sin(\theta_i)\cos(\phi_i)\sigma_x+\sin(\theta_i)\sin(\phi_i)\sigma_y+\cos(\theta_i)\sigma_z.
\end{equation}
Consequently,
\begin{eqnarray}
&&T_{(\theta_1,\phi_1)\,(\theta_2,\phi_2)\,(\theta_3,\phi_3)}\\
&&=\sin(\theta_1)\cos(\phi_1)\sin(\theta_2)\cos(\phi_2)\sin(\theta_3)\cos(\phi_3)T_{xxx}\nonumber\\
&&+\sin(\theta_1)\cos(\phi_1)\sin(\theta_2)\cos(\phi_2)\sin(\theta_3)\sin(\phi_3)T_{xxy}\nonumber\\
&&+\dots\nonumber\\
&&+\cos(\theta_1)\cos(\theta_2)\cos(\theta_3)T_{zzz}, \nonumber
\end{eqnarray}
which has to vanish since all full correlations along Pauli directions vanish.

\bibliographystyle{apsrev4-1}

\begin{thebibliography}{10}
\bibitem{footnote} {Examples of Bell inequalities involving lower order correlations can be found in~\cite{mArCIN}, however, none of them is violated by our state 
$\rho_{D_5^{(2)}}^{nc}$.}
\end{thebibliography}
%
%


\end{document}